**On the origin of neutron magnetic scattering in anti-site disordered $Sr_2FeMoO_6$ double perovskites**.

D. Sánchez, J.A. Alonso,* M. García-Hernández, M.J. Martínez-Lope, J.L. Martínez.

*Instituto de Ciencia de Materiales de Madrid, C.S.I.C., Cantoblanco, E-28049 Madrid, Spain.*

Anders Mellergård

*NFL-Studsvik, Uppsala University, S-611 82 Nyköping, Sweden*

Anti-site disordering in $Sr_2FeMoO_6$ double perovskites (containing Mo atoms at Fe positions, and viceversa) has recently been shown to have a dramatic influence in their magnetic and magnetotransport properties. In the present study, two polycrystalline $Sr_2FeMoO_6$ samples showing different degrees of anti-site disorder (a nominally "ordered" sample with ~70% of cationic ordering and a nominally "disordered" sample with ~18% of cationic ordering) have been examined by magnetic measurements and neutron powder diffraction (NPD) techniques in the 15-500K temperature range. Our main finding is that the "disordered" sample exhibits a strong magnetic scattering (noticeable even at 500K), comparable to that displayed by the "ordered" one below $T_C$= 415 K. For the "disordered" sample, the magnetic scattering exhibited on low angle Bragg positions, is not to be ascribed to a (non-existent) ferrimagnetic ordering: our results suggest that it originates upon naturally-occurring groups of Fe cations in which strong antiferromagnetic (AFM) Fe-O-Fe superexchange interactions are promoted, similar to those existing in the $LaFeO_3$ perovskite. These Fe groups are not magnetically isolated, but coupled by virtue of Fe-O-Mo AFM interactions, which maintain the long-range coherence of this AFM structure. Susceptibility measurements confirm the presence of AFM interactions below 770 K.

* Corresponding author. Electronic mail: ja.alonso@icmm.csic.es

## Introduction

Some of the oxides of the double perovskites family $A_2B'B''O_6$ (A= alkali-earth; B', B''= heterovalent transition metals) have been recently described to exhibit ferromagnetism and half-metallicity with a high spin polarization at the Fermi level, making them promising candidates as materials suitable for spin devices. The case of the half-metallic ferromagnet (ferrimagnet) $Sr_2FeMoO_6$ is paradigmatic; with a Curie temperature above room temperature ($T_C$= 415 K), it can be considered as a serious alternative to manganese perovskites for practical applications (1-5).

The ideal structure of $Sr_2FeMoO_6$ can be viewed as a regular arrangement of corner-sharing $FeO_6$ and $MoO_6$ octahedra, alternating along the three directions of the crystal, with the voluminous Sr cations occupying the voids in between the octahedra. In a simple picture, the ferrimagnetic structure can be described as an ordered array of parallel $Fe^{3+}$ (S=5/2) magnetic moments, antiferromagnetically coupled with $Mo^{5+}$ (S=1/2) spins. In this ideal model, the saturation magnetization, at low temperature, would be of 4 $\mu_B$ per formula unit (f.u.). In the real world, such a large magnetization value has not been obtained for bulk $Sr_2FeMoO_6$ up to date; instead smaller values bellow 3.7 $\mu_B$/f.u. have been reported (1,5,6). The origin of this difference with the theoretical magnetization can be found in the so-called anti-site B-cation disorder, implying that some $Mo^{5+}$ cations occupy the positions of $Fe^{3+}$ cations, and viceversa.

The problem of the order-disorder of B cations in double perovskites $A_2B'B''O_6$ is a well-known one, and it has been previously addressed (7,8). If the charge difference between B' and B'' is greater than two, complete ordering of these cations is found, for instance in perovskites of the type $A_2B^{2+}B^{6+}O_6$ and $A_2B^{+}B^{7+}O6$. For them, a perfect rock-salt like structure is obtained for the B-cations sublattice. However, for perovskites of the type $A_2B^{3+}B^{5+}O6$, various degrees of order of the B cations are observed. The actual degree of order depends mainly on synthesis conditions; it is primarily controlled by kinetic processes and not by thermodynamic equilibrium considerations. As a rule of thumb, increased order may be obtained with increased synthesis temperatures (6) or treatment time (7).

Order-disorder effects in complex oxides can significantly impact their properties. As we have briefly mentioned, the magnetization of $Sr_2FeMoO_6$ depends on the synthesis conditions, through the Fe/Mo degree of order achieved for a particular synthetic protocol (6,9,10). Moreover, we have recently demonstrated (11) that the low-field magnetoresistance of a set of $Sr_2FeMoO_6$ samples prepared under very different conditions (including soft-chemistry low-temperature procedures and high-pressure treatments) depends monotonically on the degree of B anti-site disorder of the perovskite structure.

In this paper we report on the results of a temperature-dependent neutron powder diffraction (NPD) study on two samples with very different degrees of ordering: we show that, in spite of the large difference in saturation magnetization exhibited by both samples, the magnetic neutron scattering is surprisingly similar (excepting subtle differences in its thermal evolution), when we would have expected a considerably weakened magnetic scattering in the disordered sample. We ascribe the observed scattering to Fe-rich regions, where strong Fe-O-Fe superexchange interactions give rise to the antiferromagnetic (AFM) ordering of the Fe spins.

**Experimental Section**

$Sr_2FeMoO_6$ perovskites with two different degrees of B-site ordering were prepared in pollycrystalline form by soft-chemistry procedures. Stoichiometric amounts of analytical grade $Sr(NO_3)_2$, $FeC_2O_4.H_2O$ and $(NH_4)_6Mo_7O_{24}$ . $4H_2O$ were dissolved in citric acid. The citrate + nitrate solutions were slowly evaporated, leading to an organic resin containing an homogeneous distribution of the involved cations. This resin was first dried at 120°C and then slowly decomposed at temperatures up to 600°C. All the organic materials and nitrates were eliminated in a subsequent treatment at 800°C in air, for 2 hours. This treatment gave rise to a highly reactive precursor material. The “disordered” sample was obtained after a thermal treatment at 850ºC for 2 h in a $H_2/N_2$ (15%/85%) reducing flow. The “ordered” sample was prepared from a batch of the previously synthesized “disordered” one: a fraction of this sample was retreated at 1050ºC for 12 h in an $H_2/N_2$ (5%/95%) flow, in order to favor the B-cations ordering.

The initial characterization of the products was carried out by laboratory XRD (Cu K$\alpha$, $\lambda$= 1.5406 Å). The degree of ordering of both samples was established by Rietveld analysis of the XRD patterns. NPD patterns of both samples were collected at the SLAD neutron diffractometer of the Studsvik Neutron Research Laboratory (NFL) in the temperature range 15-500K. About 5 g of sample were contained in a vanadium can; the counting time was up to 12 h for each pattern. A wavelength of 1.116 Å was used. All the patterns were refined by the FULLPROF Rietveld refinement program (12). A pseudo-Voigt function was chosen to generate the line shape of the diffraction peaks. No regions were excluded in the refinements. In the final run the following parameters were refined from the NPD data: scale factor, background coefficients, zero-point error, unit-cell parameters, pseudo-Voigt parameter, positional coordinates, isotropic thermal factors and the magnitude of the Mo an Fe ordered magnetic moments. The coherent scattering lengths for Sr, Fe, Mo and O were, 7.02, 9.45, 6.72 and 5.803 *fm,* respectively. The *dc* magnetic susceptibility was measured with a commercial SQUID magnetometer on powdered samples, in the temperature range 5 to 450 K.

## Results

$Sr_2FeMoO_6$ oxides were obtained as black, well crystallized powders. The laboratory XRD diagrams at RT are shown in Fig. 1. The patterns are characteristic of a perovskite structure; the “ordered” sample shows superstructure peaks arising from the Fe/Mo ordering (e.g. (011) and (013)), which are absent in the “disordered” sample. From the analysis of the intensities of these reflections, via Rietveld refinements of the XRD data, the degree of ordering was estimated to be of 68% and 18% for the “ordered” and “disordered” samples, respectively. If we define the parameter $x$ as the fraction of Mo atoms at Fe positions, we have $x$=0.41 for the “disordered” sample and $x$=0.16 for the “ordered” sample. Notice that $x$ would take a value of 0.5 for a completely disordered sample.

## Magnetic data

The magnetization *vs.* temperature data of the “ordered” sample (Fig. 2a) show a low temperature saturation characteristic of a spontaneous ferromagnetic (FM) ordering. The magnetization *vs.* magnetic field data shown in Fig. 3a at 5 K are characteristic of a ferromagnet with a saturation magnetic moment of 2.8 $\mu_B$/f.u. Taking the first derivative of the M(T) curve we stablish a $T_c$ of 415 K. By contrast, the data for the "disordered" perovskite (Figs. 2a, 2b and 3b) show a more progressive evolution (almost linear) of the susceptibility as temperature decreases, reaching a much weaker saturation moment at 5K, of only 0.8 $\mu_B$/f.u. It is striking the presence, in the high-temperature region, of a well-defined maximum centered at 770 K (Fig. 2b), which suggests the establishment of AFM interactions below this temperature.

The strong suppression of the FM properties in the "disordered" sample is due to the anti-site Fe/Mo disorder in the B positions of the perovskite, if we consider this saturation magnetization ($M_s$) to be the result of an ideal ferrimagnetic ordering between the moments of the B' and B'' positions of the double perovskite, as $M_s= M_{B'}-M_{B''}$. In a simple picture, assuming localized magnetic moments of 5$\mu_B$ for $Fe^{3+}$ and 1$\mu_B$ for $Mo^{5+}$, the anti-site occupancy of Mo at Fe positions and viceversa would give a variation of the saturation magnetization as $M_s$=(4- 8*x*) $\mu_B$/f.u. (6,10), where *x* is the fraction of Fe atoms replaced by Mo. In our ordered and disordered samples, the calculated $M_s$ values for the observed degree of ordering are 2.72 and 0.72 $\mu_B$ /f.u respectively, in good agreement with the observed $M_s$ values.

## Structural refinement

Above 415K the structure of the "ordered" sample was refined in the cubic Fm3m space group (No. 225), with a doubled unit-cell parameter **c**≈ 2$a_0$, $\mathbf{a_0}$≈ 3.9 Å. Table I includes the unit-cell, atomic and thermal parameters, and discrepancy factors after the refinement at 500 K. Fig. 4a shows the goodness of the fit for the 500 K pattern. Below 415 K the crystal structure was defined in the I4/m space group (No. 87), Z=2, with unit-cell parameters related to $\mathbf{a_0}$ (ideal cubic perovskite, $\mathbf{a_0}$≈ 3.9 Å) as **a**= **b** ≈ √2$\mathbf{a_0}$, **c**≈ 2$a_0$. Sr atoms were located at *4d* positions, Fe at *2a*, Mo at *2b* sites, and oxygen atoms at *4e* and *8h* positions. Fig. 4b illustrates the profile agreement for the 15K pattern, including the magnetic structure refinement which is described below. The structural results for the "ordered" sample completely agree with those recently published by Chmaissen et al. (13).

For the “disordered” sample the structure was found to be tetragonal in all the temperature range, from 15 to 500K; thus all of the NPD patterns were refined in the I4/m space group. The anti-site disordering obtained from the XRD patterns was included in the model, but it was not refined since a strong correlation with the magnetic contribution to the scattering was found in trial refinements. The unit cell parameters and volume variation with temperature for both samples is represented in Fig. 5, as well as the thermal evolution of the tetragonal strain, defined as $c-\sqrt{2}a$. A larger volume is observed for the disordered sample, in which the electrostatic repulsion of the highly charged $Mo^{5+}$ cations is not minimized due to the frequent occurrence of Mo-O-Mo configurations.

**Magnetic structures**

For both ordered and disordered samples, the low-temperature NPD data reveal a strong magnetic contribution on the low angle reflections. The magnetic scattering can be mostly appreciated on the (011) and (013) superstructure reflections; the evolution of both reflections at selected temperatures is displayed in Fig. 6. The thermal variation of the integrated intensity for the (011) reflection is plotted in Fig. 7 for the ordered and disordered samples. We considered two different models to describe the magnetic structures of both samples.

*Ordered sample*

In a first trial, a FM structure was modeled with magnetic moments only at the Fe positions; after the full refinement of the profile for the 15 K NPD data, including the magnetic moment magnitude and orientation, a discrepancy $R_{mag}$ factors of ~8% was reached. The subsequent introduction of magnetic moments at the Mo positions in an AFM arrangement with respect to Fe moments (i.e. describing a global ferrimagnetic structure) lead to a dramatic improvement of the refinement: $R_{mag}$ dropped to ~4% for the final parameters listed in Table I. After the final refinement, ordered moments of 3.9(1) $\mu_B$ and -0.37(6) $\mu_B$ were obtained for Fe and Mo positions, respectively. Trials to couple ferromagnetically Fe and Mo moments invariably led to a serious deterioration of the fit. The spatial orientation of the moments is affected by a large error and was not refined; a fixed orientation parallel to the c axis was considered. The same model for the magnetic structure was used to refine the remaining NPD patterns; the parameters were refined sequentially on increasing temperature, up to 415 K. Fig. 8 shows the thermal evolution of the ordered magnetic moment at Fe positions.

*Disordered sample*

The most surprising result of this work was the observation of a strong magnetic contribution to the scattering in the disordered sample, for which the saturation magnetization is very small, hardly 0.8 $\mu_B$ /f.u. A priori we would have expected that the structural disordering at the B position of the perovskite would have led to the absence of long-range magnetic ordering between Fe and Mo magnetic moments. However, in Fig. 6 we observe a magnetic contribution, similar in magnitude, on the same low-angle reflections for both samples. Surprisingly, the integrated intensity of the (011) reflections are quite comparable at 15 K (Fig. 7); moreover it is slightly larger in the “disordered” sample in all the temperature range. An abrupt decay of the intensity is observed above $T_C$ for the “ordered” sample, as expected; the residual intensity corresponds to the structural Fe/Mo ordering. As for the disordered sample, the decay is more gradual and a substantial magnetic contribution is still present at 500 K, as it can also be observed in Fig. 6.

As it will be discussed later, we believe that the magnetic scattering of the disordered sample originates from AFM coupling between neighboring Fe cations through Fe-O-Fe paths, which naturally occur in a random distribution of Fe/Mo cations. We have modeled, thus, the magnetic structure as a perfect AFM arrangement of Fe spins with alternating directions, occupying all the B positions of a perovskite structure with the same unit cell and positional parameters as the host $Sr_2FeMoO_6$ perovskite. We have constrained the scale factor of both structural and magnetic models, and refined the magnitude of the Fe magnetic moments. The parameters after the Rietveld refinement of the 15 K pattern are included in Table I. The goodness of the fit is shown in Fig. 4d, where both structural and magnetic contributions to the scattering have been included.

The thermal variation of the magnitude of the Fe magnetic moments is included in Fig. 8.

Notice that the ideal AFM model for the magnetic structure of the “disordered” sample implies the absence of magnetic scattering at the (h,k,l) peaks when l= even, whereas for the ferrimagnetic structure of the “ordered” sample the magnetic contribution occurs at both l= even and odd. This is shown in Fig. 9, where the magnetic contributions to the scattering for both samples are presented. This effect (the absence of magnetic peaks for l= even) is impossible to be observed directly from the diffraction patterns, as there is always a strong crystallographic contribution on the l=even Bragg positions. Only the Rietveld refinement of both patterns allowed us to confirm the proposed model. Trials to refine the “disordered” structure with the ferrimagnetic model (starting from 3.9 $\mu_B$ at Fe positions and 0.37 $\mu_B$ at Mo positions) led to a fast increase of the magnetic moment at the Mo positions, (and a decrease at Fe positions), leading to a model close to that proposed for the “disordered” sample, with comparable moments at all positions. In fact, a slightly larger moment was observed on Fe positions, which is due to the contribution of the 18% of Fe/Mo ordering present in the sample.

It is also worth underlying that the presence of a FM coupling between near neighbor Fe-Fe magnetic moments would inevitably imply the absence of any magnetic intensity on the (011), (013)… Bragg positions, in contradiction with the observation.

**Discussion**

A tetragonal to cubic structural transition concomitant with the FM transition ($T_C$) has been recently described for $Sr_2FeMoO_6$ (13,14), which is in agreement with our observations. It is commonly admitted that the high-temperature structure (above $T_C$) crystallizes in the Fm3m space group, which allows for a rock-salt like distribution of Fe and Mo cations over the B sublattice positions, in perfectly ordered samples, and implies a Fe-O-Mo angle of 180º. However, there is still some discrepancy regarding the space group of the tetragonal low-temperature phase (below $T_C$); although the space groups I4/mmm and P4/mmm (6, 14) have been suggested, we have adopted the model recently described by Chmaissen et al (13), in I4/m . The structure of $Sr_2FeMoO_6$ can be described as the result of a single anti-phase octahedral tilting along the **c**-axis. The magnitude of the tilting can be simply derived from the Fe-O2-Mo angle; this angle (Fig. 10) evolves from a maximum value at 15 K ($\phi$= 5.5º for the "ordered" sample) to $\phi$= 0 at $T_C$, at the onset of the structural phase transition from tetragonal (low temperature) to cubic (high temperature). For the "disordered" sample the structure is still tetragonal at the maximum measurement temperature of 500 K, at which the tilting angle takes a significant value of $\phi$= 2.5º.

As shown in Table II, in the "ordered" sample $FeO_6$ octahedra are significantly larger (expanded) than $MoO_6$ octahedra. This observation is coherent with the larger ionic size of $Fe^{3+}$ *vs.* $Mo^{5+}$(15). For the "disordered" sample the Fe-O and Mo-O bond lengths are more similar, as expected for the high degree of anti-site disordering.

Although the actual electronic configurations $Fe^{3+}(3d^5)$-$Mo^{5+}(4d^1)$ *vs.* $Fe^{2+}(3d^4)$-$Mo^{6+}(4d^0)$ have been considered as possible, the average valence for Fe has been found to be intermediate between high-spin configuration values of $Fe^{2+}$ and $Fe^{3+}$ from Mössbauer spectroscopy studies (16). This is to say that both electronic configurations (with $Fe^{2+}$ and $Fe^{3+}$) must be considered as degenerate, the final state being a combination of both configurations. At 15 K we find a small but significant ordered magnetic moment on the Mo sites ($\mu_{Mo}$= 0.37 ±0.06 $\mu_B$), antiferromagnetically coupled with the Fe magnetic moments in a ferrimagnetic configuration, which is clearly consistent with an intermediate V-VI oxidation state for Mo.

The main issue to be addressed in this paper is the origin of the magnetic scattering on the sample showing an important component of anti-site disordering. We have been able to accurately fit the magnetic contributions to the neutron scattering by modeling an AFM structure consisting of a perfect arrangement of Fe cations occupying all of the B positions of a perovskite structure with the same unit-cell parameters as the crystallographic $Sr_2FeMoO_6$ phase. The refinement of the magnitude of the magnetic moments on the Fe positions, at 15 K, gives an ordered magnetic moment of 2.2 $\mu_B$, with constrained scale factors for the crystal and magnetic structure. An equivalent approach is to constrain the $Fe^{3+}$ magnetic moment to 5 $\mu_B$ and to refine the scale factor of the magnetically diffracting phase: by doing so we have obtained that about one half of the main crystallographic phase is also magnetically diffracting. This is to say that, in spite of the disordered nature of the Fe distribution upon the crystal structure, almost all of the $Fe^{3+}$ cations are actively participating in the magnetic scattering.

The almost random distribution of Fe/Mo cations that we expect for the “disordered” sample implies that the Fe-O-Fe configuration frequently occur (as well as Mo-O-Mo configurations); in these regions we must consider that the Fe-O-Fe magnetic interactions are by far more similar to those happening in $LaFeO_3$ perovskite than in $SrFeO_3$. The former $Fe^{3+}$-containing perovskite experiences an AFM ordering below a surprisingly high Néel temperature, $T_N$= 750 K (17). However, for the $Fe^{4+}$-containing $SrFeO_3$ perovskite, $T_N$ is much lower, of 134 K (18). Our high-temperature susceptibility measurements (Fig. 2b) show a maximum at 770 K, suggesting AFM interactions of similar strength to those observed in $LaFeO_3$. In the regions where the Fe-O-Fe configurations occur, in spite of having a local chemical composition closer to $SrFeO_3$, we must not forget that the electronic delocalization of the Mo-electron that is still present despite the B-cation disordering enables the establishment of a global charge neutrality across the crystal, keeping the nominally trivalent valence for Fe.

Our picture thus shows a disordered Fe/Mo pattern in which the Fe-O-Fe superexchange AFM interactions are comparable to those existing in $LaFeO_3$. Fig. 11 shows a simplified image of the magnetic ordering in the disordered sample; only one layer of B cations is shown for the sake of simplicity. The coupling between near-neighbor Fe-Fe or Fe-Mo atoms is always AFM (as it happens in the perfectly ordered perovskite). The important fact is that the coherence between the magnetic ordering of isolated Fe-O-Fe couples is maintained by the intermediate Mo atoms, by virtue of the also AFM nature of the Fe-O-Mo interactions. We cannot argue if isolated Mo atoms inside Mo-O-Mo regions (without Mo-Fe contacts) are magnetically ordered; in fact this is irrelevant for our model of magnetic structure, in which we only consider the antiferromagnetically ordered Fe-O-Fe regions, mostly contributing to the magnetic scattering. The small extra contribution of the Mo-rich regions can be neglected for our purposes.

The thermal variation of the magnetic moment on Fe positions (Fig. 8) shows a distinct behavior for both samples. The Fe moment for the “ordered” perovskite exhibits an abrupt decay at 415 K, which corresponds to the vanishing of ferrimagnetic ordering, whereas the Fe moment for the “disordered” sample gradually decreases and it is still significant at the highest measurement temperature, of 500 K. This behavior can be understood on the basis of the stronger superexchange Fe-O-Fe interactions operating in the “disordered” sample. An extrapolation of the data represented in Fig. 8 gives a Néel temperature for the “disordered” sample around 760 K, close to that the maximum observed in the susceptibility curve (Fig. 2b). At the same time, while the “ordered” sample exhibits a phase transition from tetragonal to cubic upon heating across $T_C$, the “disordered” perovskite is still tetragonal at the highest measurement temperature of 500K, as clearly suggested by the unit-cell parameters and strain thermal variation, in Fig. 5. This fact could explain the controversy observed in related literature, in which the description of $Sr_2FeMoO_6$ samples has been assigned to both cubic and tetragonal symmetries at RT, probably depending on the degree of ordering of these particular samples.

It is worth commenting that recent ab-initio calculations of the disorder effects on electronic structure and magnetic structure suggest (19) that in the disordered samples the individual magnetic moments at each Fe site are strongly reduced due to the destruction of the metallic state, while the coupling between the various Fe sites continue to be FM: their results suggest that near-neighbor Fe-Fe and Mo-Mo interactions are FM in origin. This is in contrast with our experimental observation; moreover, the presence of a strong magnetic contribution to the scattering on the (011) superstructure reflection clearly indicates that the FM alignment of near-neighbor atoms is to be excluded: only a AFM arrangement of the corresponding spins gives rise to a magnetic superstructure and explains the diffraction on these Bragg positions.

In agreement with our findings, Montecarlo calculations (10) predicted that disorder may lead to AFM couplings between neighboring Fe sites, instead of the FM coupling proposed for the idealized structure. Furthermore, the possibility of existence of strong AFM couplings between adjacent Fe-Fe next-neighbors has also been suggested by Dass and Goodenough (20), who attribute the little hysteresis exhibited by the M/H curves of $Sr_2FeMoO_6$ to AFM Fe-O-Fe interactions across antiphase boundaries. Moreover, the anomalous susceptibility and the abnormal increase of the paramagnetic Weiss constant can be accounted for (20) by a chemical inhomogeneity in disordered samples consisting of a layer of three (111) all-Fe planes giving rise to strong 180º Fe-O-Fe superexchange interactions between the layers.

## Conclusions

Neutron diffraction data on a disordered $Sr_2FeMoO_6$ sample shows a strong magnetic contribution on superstructure reflections, which provides strong experimental support to the hypothesis that near-neighbor Fe-Fe atoms are antiferromagnetically coupled, by virtue of strong superexchange Fe-O-Fe interactions. In a pattern of randomly distributed Fe and Mo atoms, the magnetic scattering originates from naturally occurring Fe-O-Fe pairs, experiencing AFM Fe-Fe interactions; the refinement of the magnitude of the ordered magnetic moment shows that virtually all of the Fe atoms participate in the scattering process. The coherence of the AFM arrangement of neighboring Fe-O-Fe regions is maintained across the crystal by AFM Fe-Mo superexchange interactions. The chemical disorder, giving rise to Fe-O-Fe and Mo-O-Mo couples, seems to be accompanied by a charge segregation in such a way that an average trivalent oxidation state is attributable to Fe atoms in Fe-rich regions. Given the different origin of the main magnetic interactions with respect to the ordered sample, the thermal variation of the ordered magnetic moments in the "disordered" sample shows a slower decay upon heating; this sample is antiferromagnetically ordered below 770 K.

## Acknowledgements

We thank the financial support of CICyT to the projects PB97-1181, MAT99-1045 and the European Community - Access to Research Infrastructure action of the Improving Human Potential Programme (HPRI-CT-1999-00061) for making feasible our access to the NFL-Studsvik. We deeply acknowledge to María Martínez her help during the experiment.

**Figure captions**

Fig. 1. XRD pattern (Cu Kα) for $Sr_2FeMoO_6$ perovskites at RT. The insets show the first two superstructure peaks ((011) and (013)) in the tetragonal setting, with **a**= **b** ≈√2$\mathbf{a_0}$, **c**≈ 2$a_0$, $\mathbf{a}_0$≈ 3.9 Å) for the ordered and disordered samples studied in the present work.

Fig. 2. Magnetization at 5000 Oe for (a) ordered and (b) disordered perovskites.

Fig. 3. Magnetization *vs.* field isotherms (T= 5K) for (a) ordered and (b) disordered perovskites.

Fig. 4. Observed (open circles), calculated (full line) and difference (bottom) NPD Rietveld profiles for $Sr_2FeMoO_6$ perovskites: (a) ordered, 500 K (Fm3m), (b) ordered 15 K (I4/m), (c) disordered 500 K (I4/m), (d) disordered 15 K (I4/m).The second series of tick marks correspond to the magnetic Bragg reflections.

Fig. 5. Thermal variation of (a) unit-cell parameters, (b) unit-cell volume and (d) tetragonal strain, defined as s= c-√2a.

Fig. 6. Low-angle region of the NPD patterns, (a) ordered sample and (b) disordered sample, showing the evolution of the magnetic scattering on the (011) and (013) superstructure reflections.

Fig. 7. Temperature evolution of the integrated intensity of (011) reflection for the "ordered" and "disordered" patterns.

Fig. 8. Thermal dependence of the Fe magnetic moments for the "ordered" and "disordered" perovskites.

Fig. 9. Observed NPD patterns (open circles) and calculated magnetic scattering (thick line) for a ferrimagnetic model (ordered sample) and AFM model (disordered sample). The magnetic scattering on (hkl), l=even reflections occurs only for the ferrimagnetic model.

Fig. 10. Temperature dependence of the tilting angle of the $FeO_6$ and $MoO_6$ octahedra.

Fig. 11. Schematic view of the magnetic coupling in a disordered sample, showing a high-degree of B-antisite disordering.

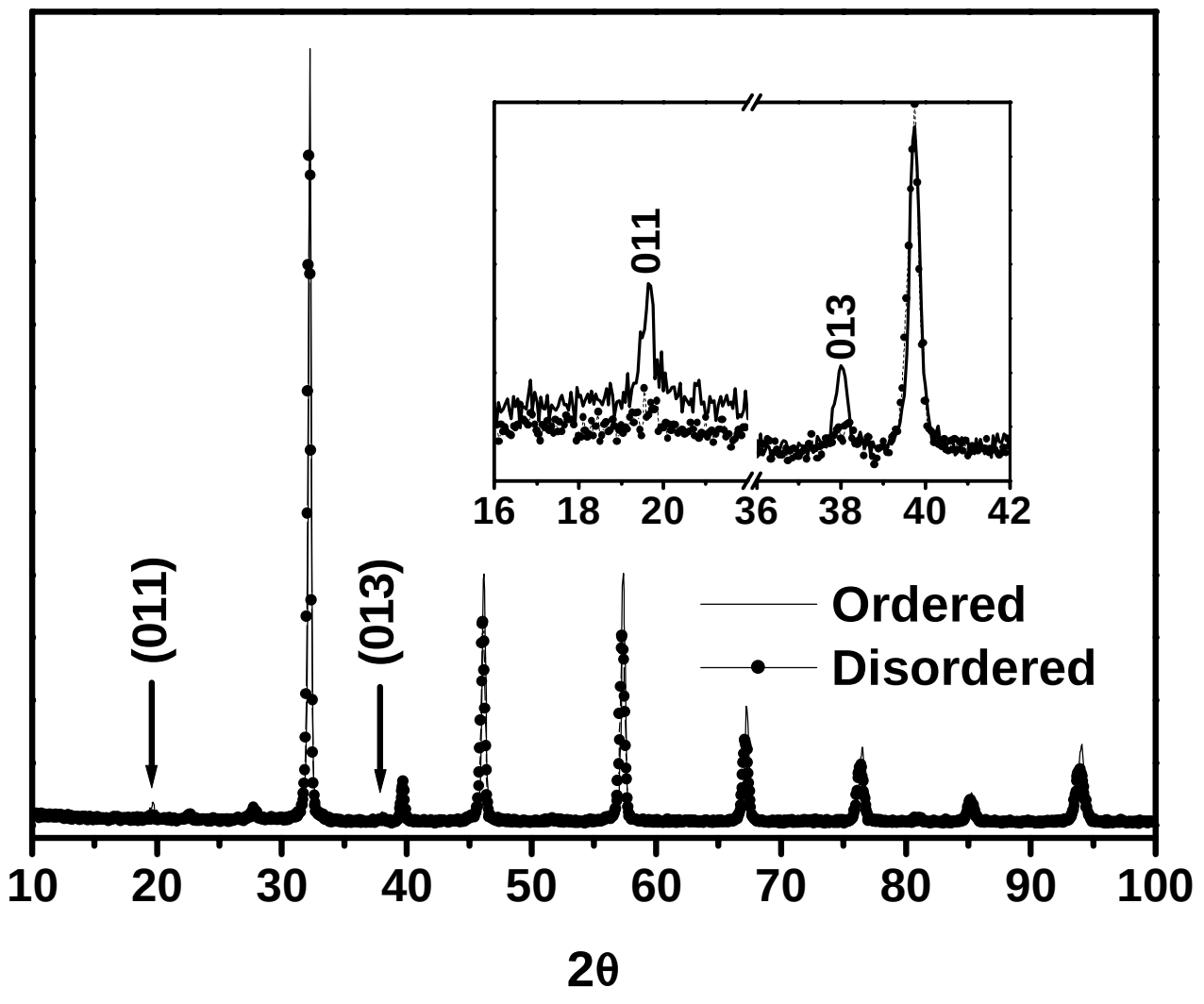

**Fig. 1**

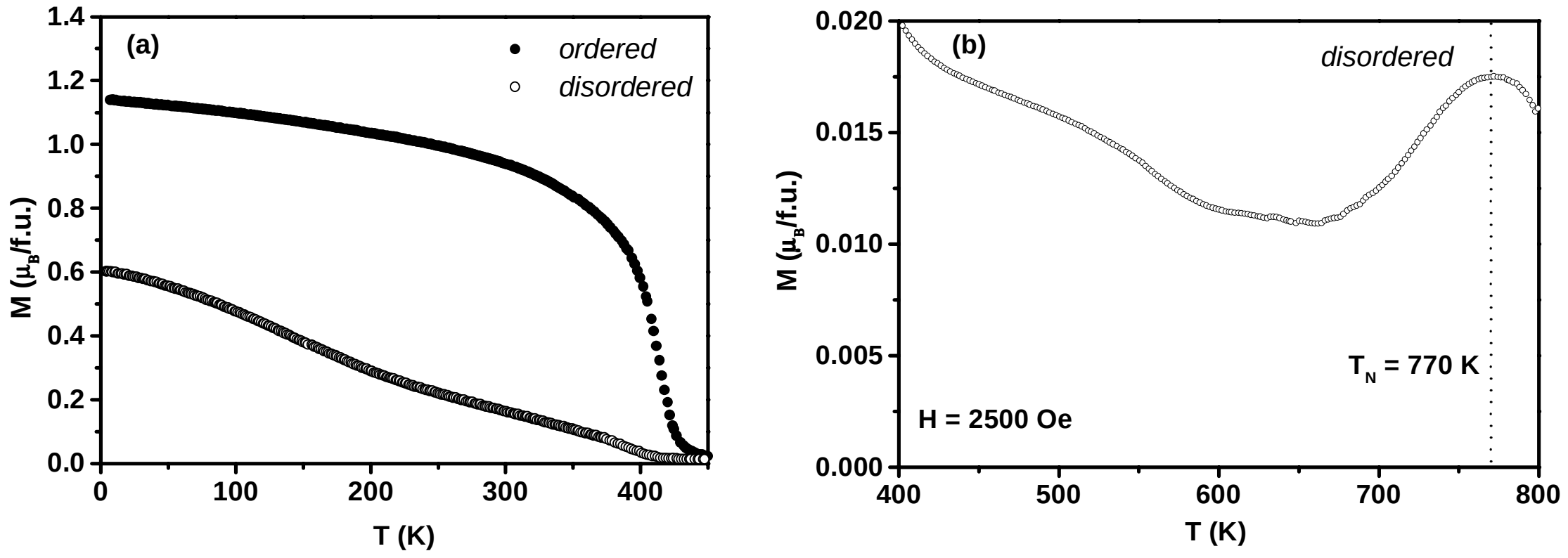

**Fig. 2**

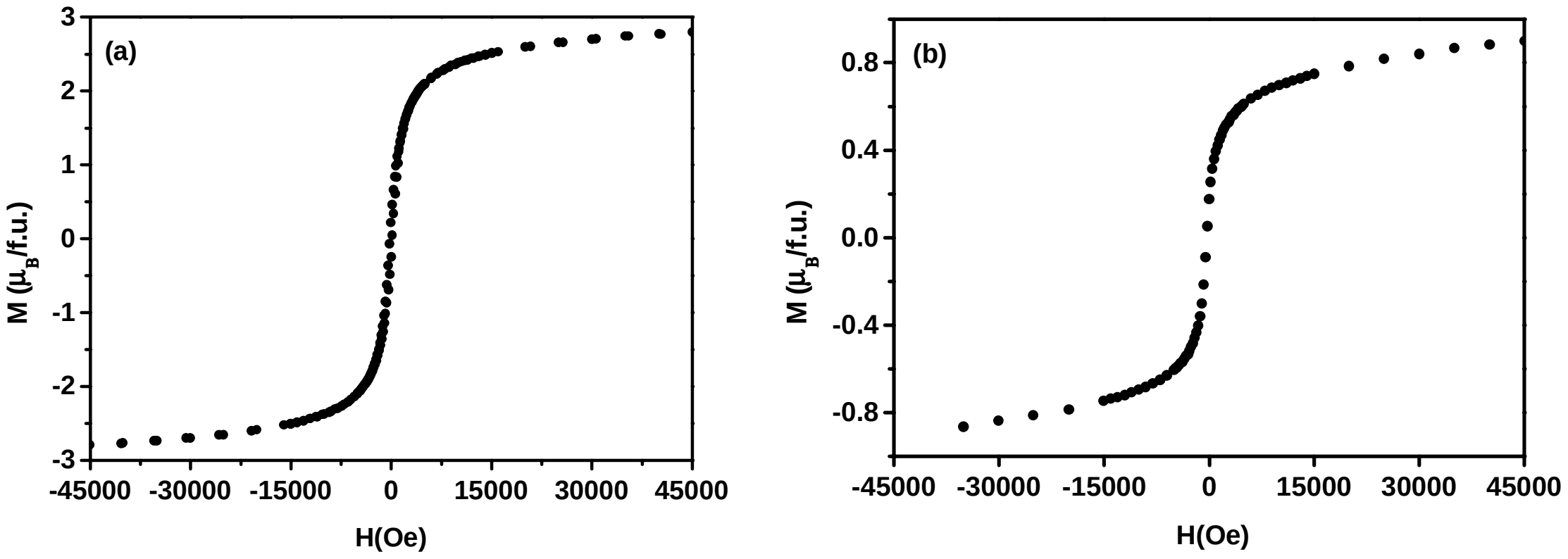

**Fig. 3**

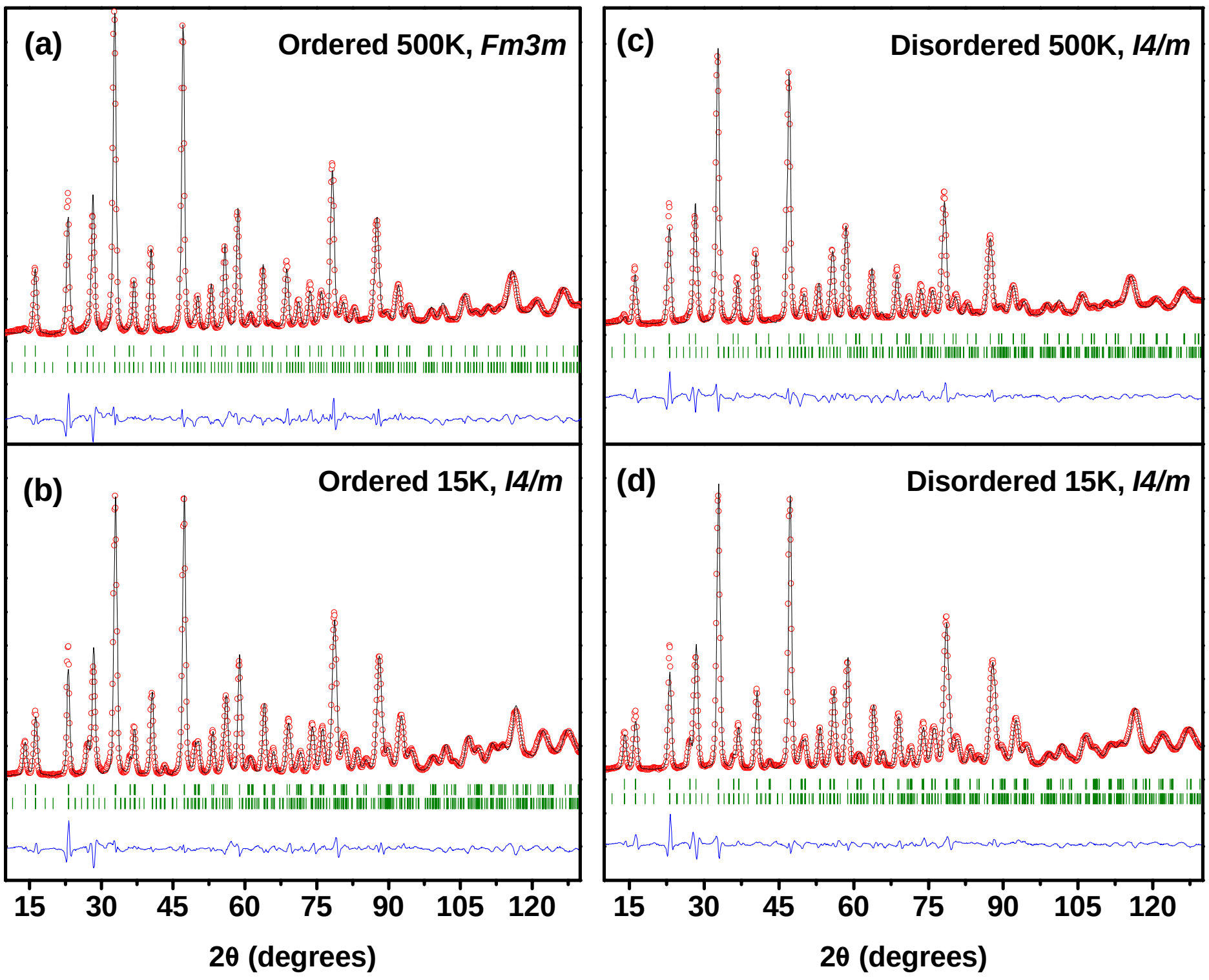
(a)
Ordered 500K, *Fm3m*
(b)
Ordered 15K, *I4/m*
(c)
Disordered 500K, *I4/m*
(d)
Disordered 15K, *I4/m*
15 30 45 60 75 90 105 120
2θ (degrees)
15 30 45 60 75 90 105 120
2θ (degrees)

Fig. 4

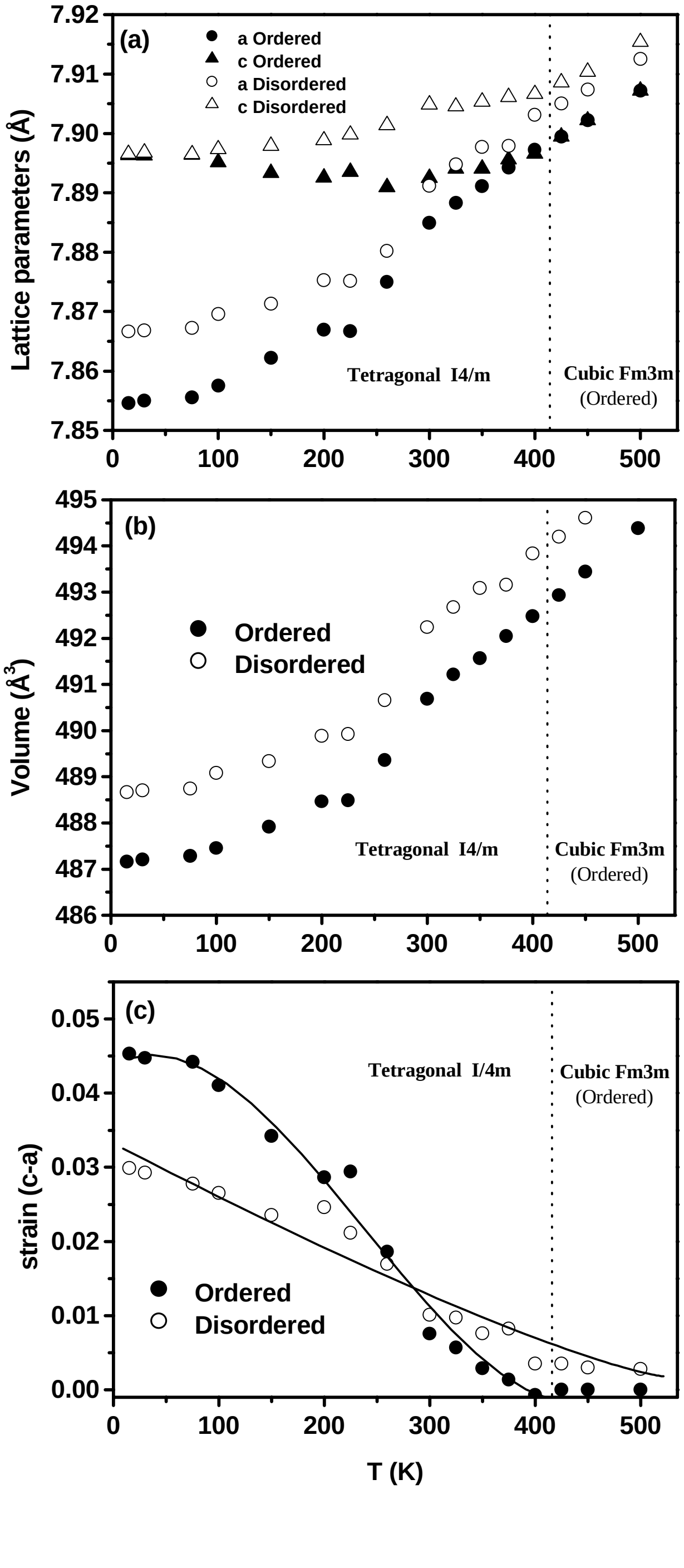

**Fig. 5**

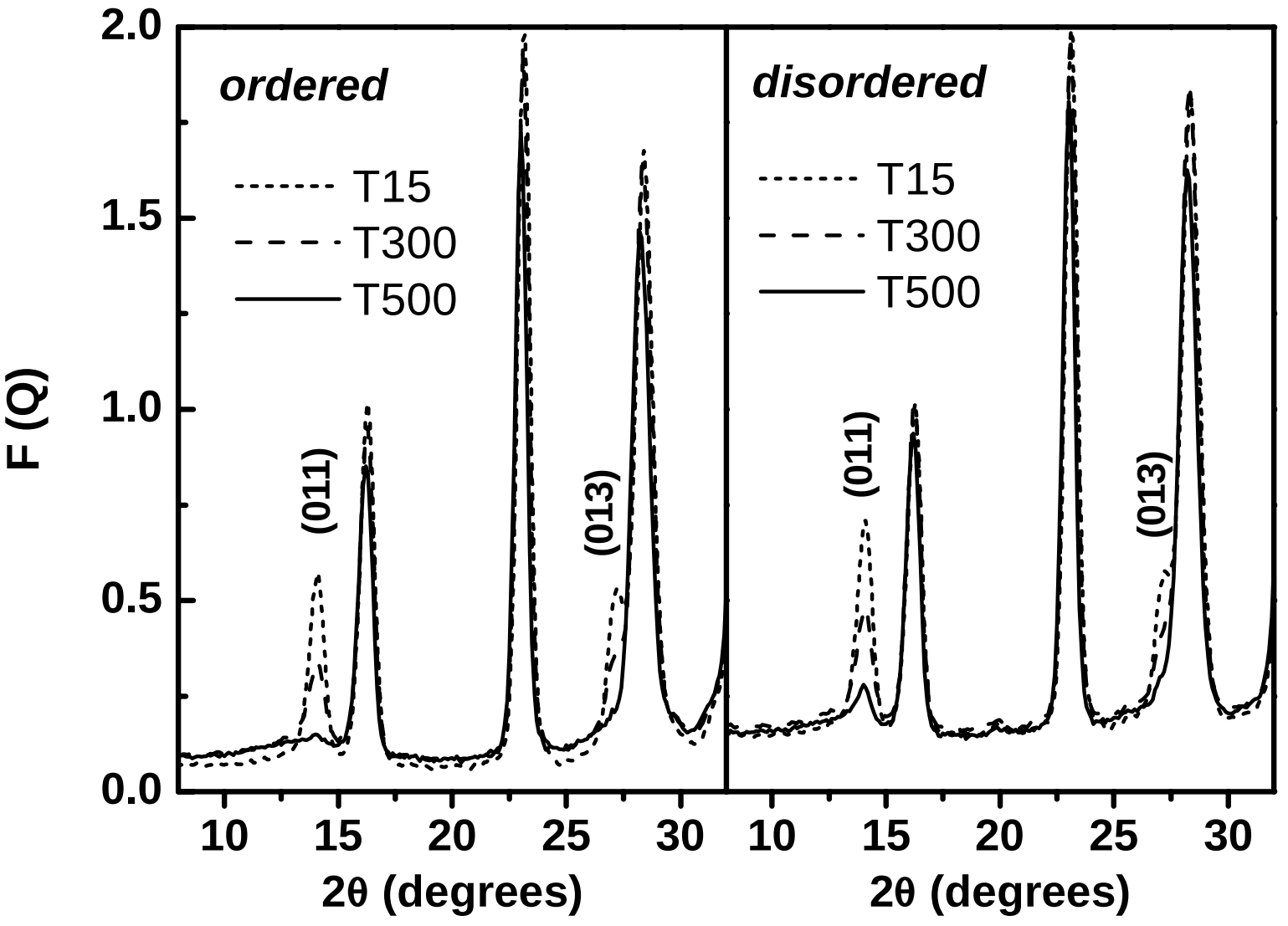

**Fig. 6**

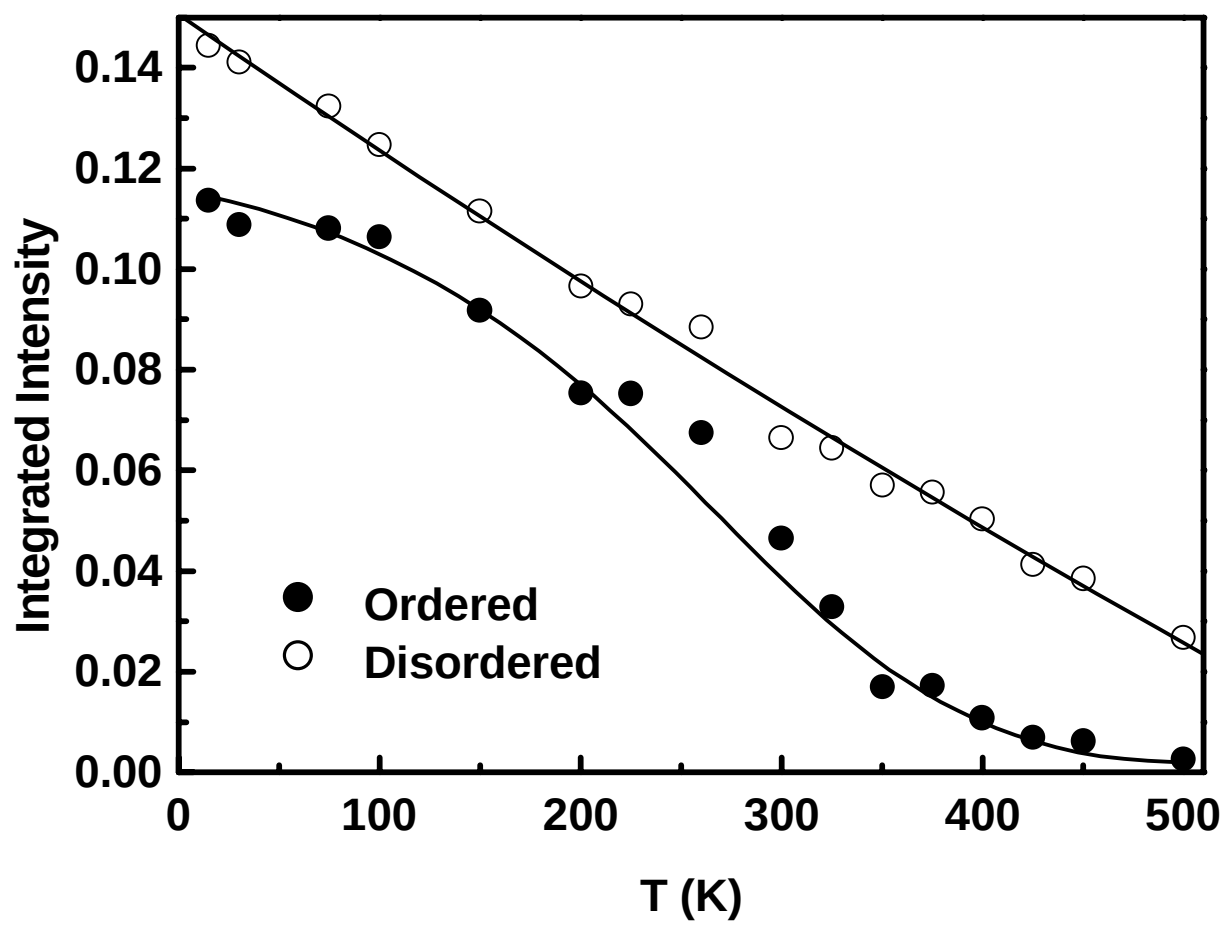

**Fig. 7**

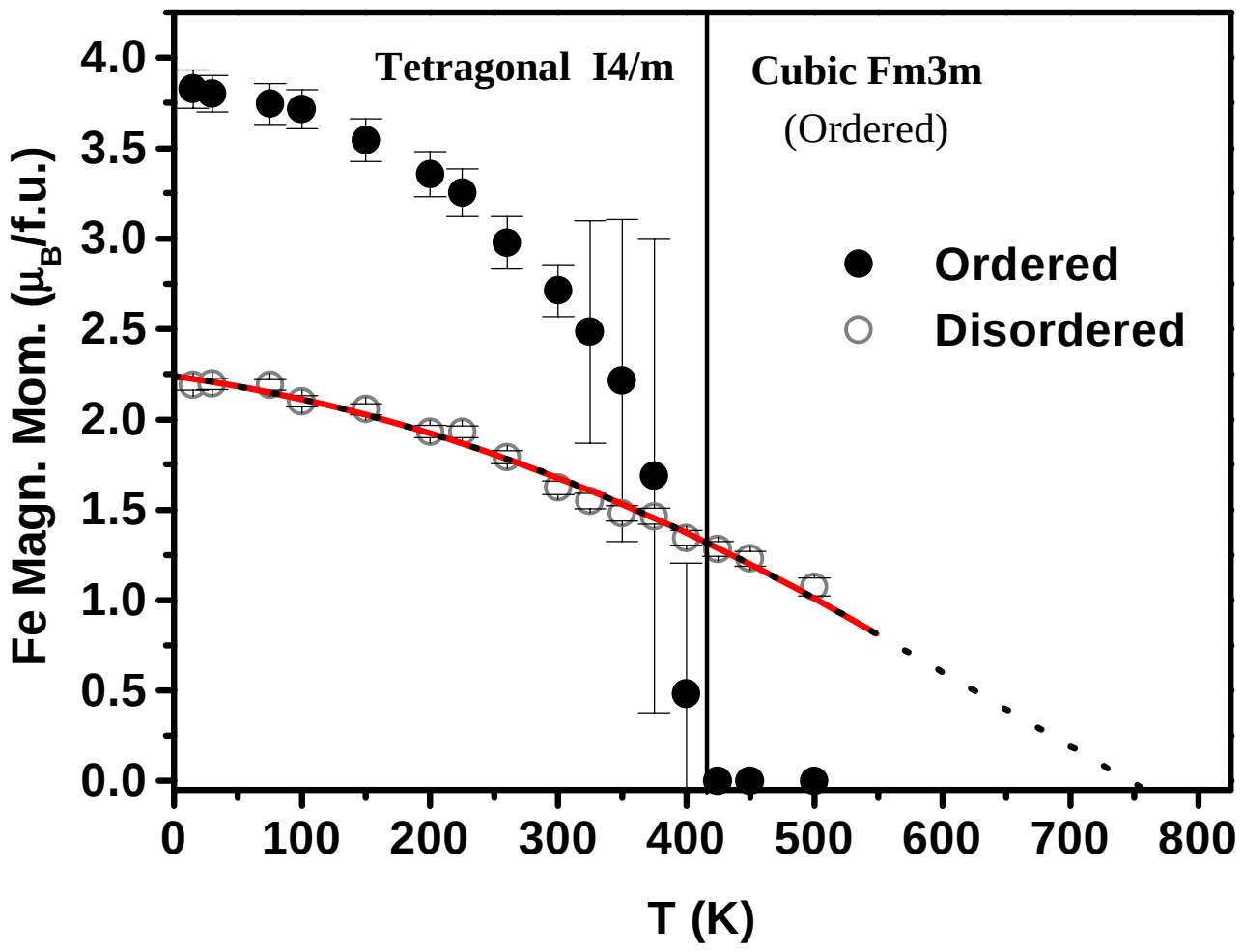

**Fig. 8**

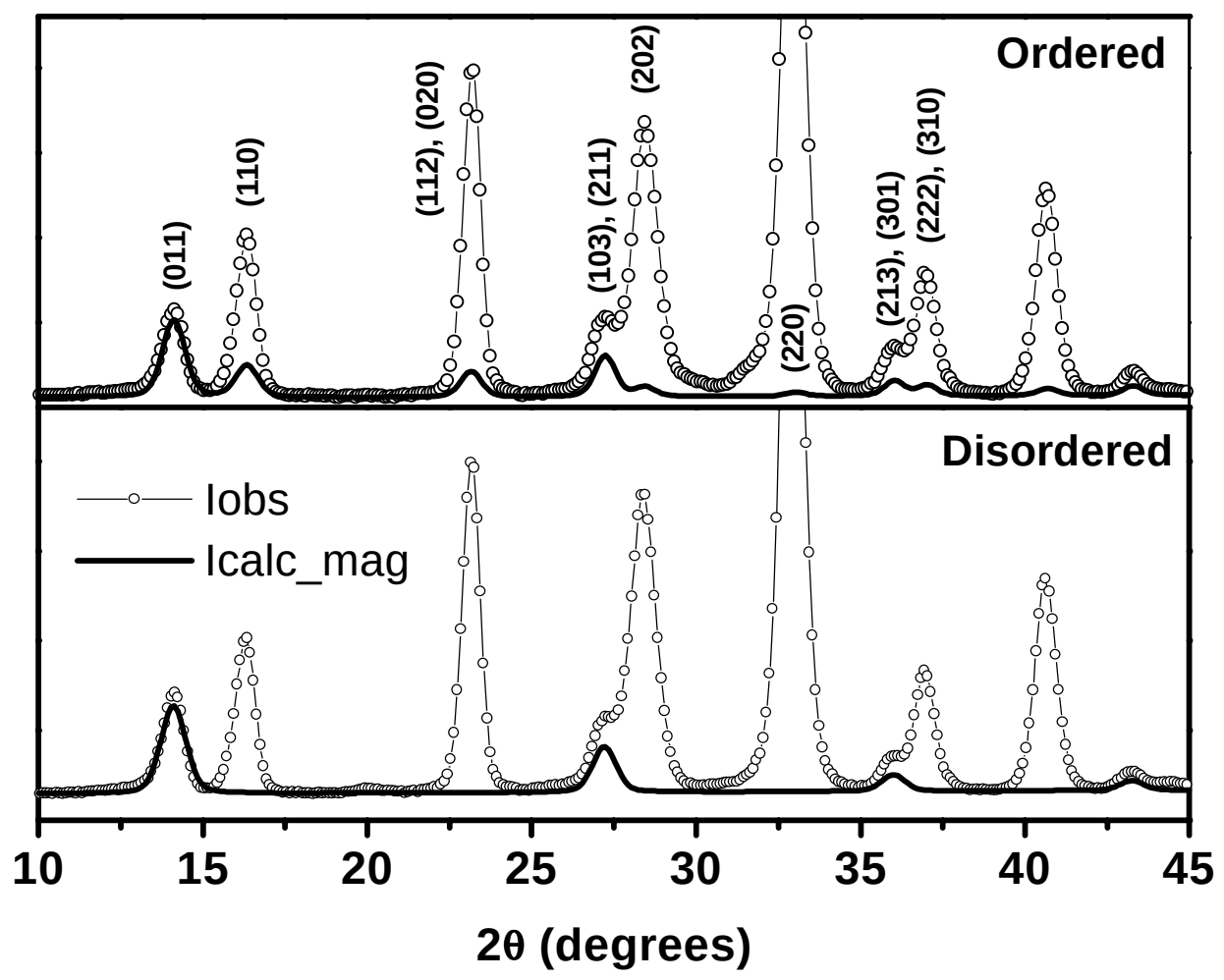

**Fig. 9**

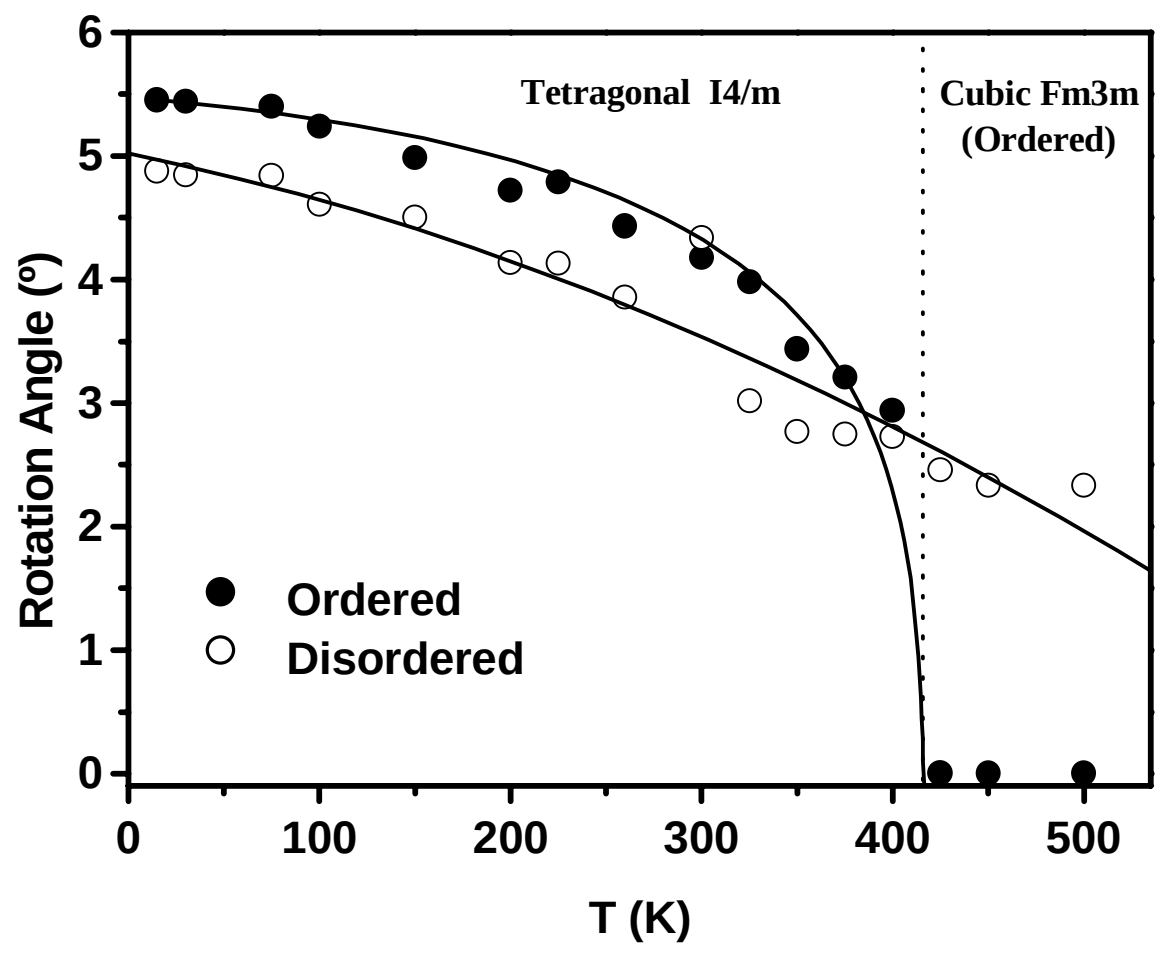

**Fig. 10**

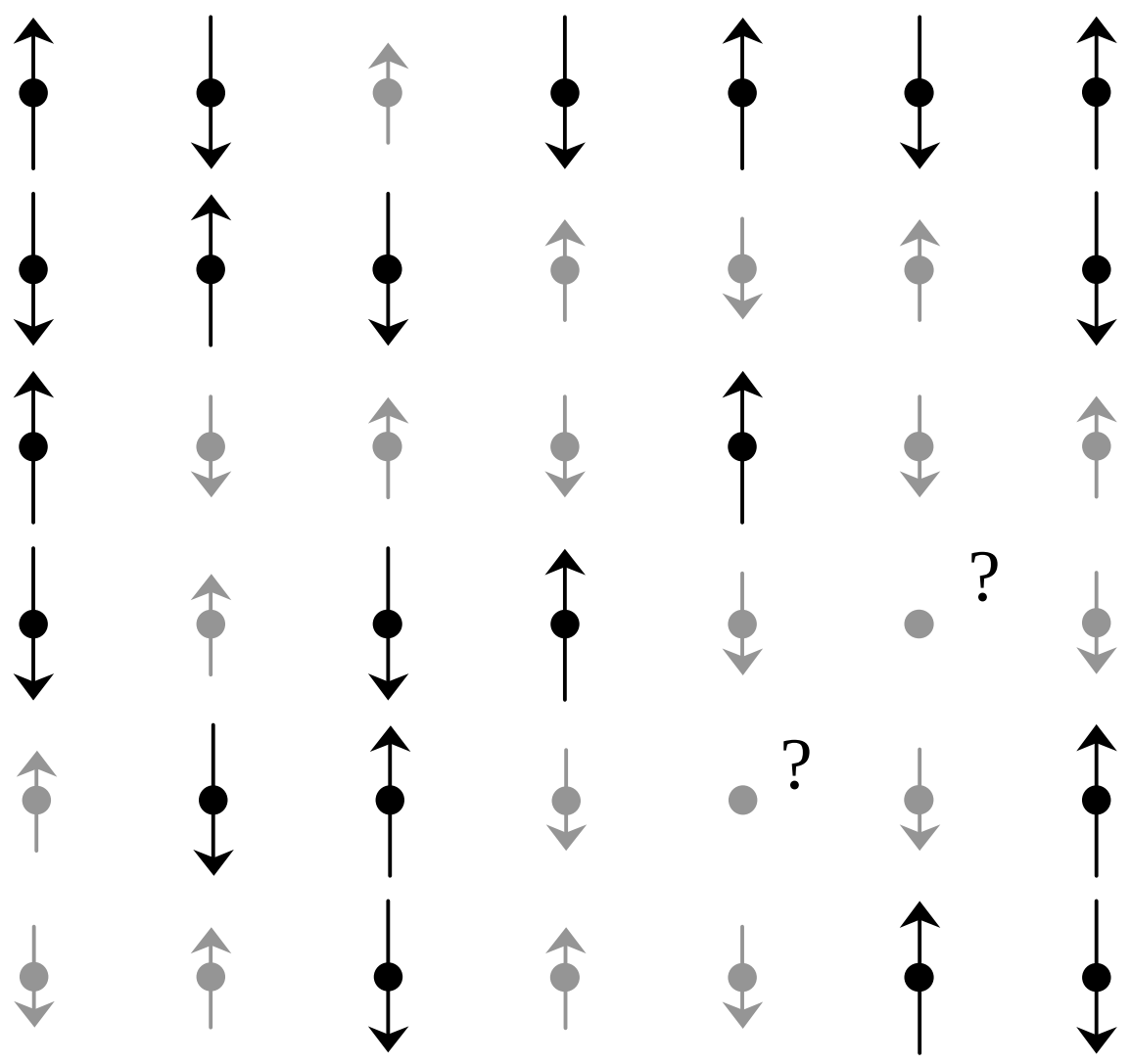

**Fig. 11**

**Table I.-** Atomic parameters after the Rietveld refinement of NPD patterns for ordered and disordered $Sr_2FeMoO_6$ perovskites

| | 15 K | | 500 K | | |
|---|---|---|---|---|---|
| Sample | Disordered | Ordered | Disordered | | Ordered |
| *I4/m* | | | | *Fm3m* | |
| a(Å) | 5.56241(7) | 5.5540(5) | 5.5950(3) | a(Å) | 7.9072(4) |
| c(Å) | 7.8969(2) | 7.900(1) | 7.9154(8) | V($Å^3$) | 494.39(8) |
| V($Å^3$) | 488.67(2) | 487.4(1) | 495.6(1) | | |
| Sr 4d($^1/_2$ 0 $^1/_4$) | | | | Sr 8c($^1/_4$ $^1/_4$ $^1/_4$) | |
| B($Å^2$) | 0.31(2) | 0.20(2) | 1.05(3) | B($Å^2$) | 1.05(3) |
| Fe 2a(0 0 0) | | | | Fe 4a(0 0 0) | |
| B($Å^2$) | 0.29(2) | 0.2(1) | 0.73(3) | B($Å^2$) | 0.6(1) |
| Magn. mom. ($\mu$B) | 2.20(3) | 3.9(1) | 1.02(6) | Magn. mom. ($\mu$B) | - |
| Mo 2b(0 0 $^1/_2$) | | | | Mo 4b($^1/_2$ $^1/_2$ $^1/_2$) | |
| B($Å^2$) | 0.01(2) | 0.1(2) | 0.43(3) | B($Å^2$) | 0.3(1) |
| Magn. mom. ($\mu$B) | - | -0.37(6) | - | Magn. mom. ($\mu$B) | - |
| O1 4e(0 0 z) | | | | O 24e(x 0 0) | |
| z | 0.249(3) | 0.254(4) | 0.256(2) | x | 0.251(1) |
| B($Å^2$) | 0.61(5) | 0.55(8) | 0.4(1) | B($Å^2$) | 1.28(2) |
| O2 8h(x y 0) | | | | | |
| x | 0.272(3) | 0.277(2) | 0.257(2) | | |
| y | 0.227(3) | 0.228(2) | 0.232(2) | | |
| B($Å^2$) | 0.42(3) | 0.25(5) | 1.6(1) | | |
| Reliability factors | | | | Reliability factors | |
| $\chi^2$ | 10.4 | 10.9 | 12.0 | $\chi^2$ | 7.93 |
| $R_p$ (%) | 7.23 | 7.94 | 11.1 | $R_p$ (%) | 11.8 |
| $R_{wp}$ (%) | 9.43 | 9.42 | 13.0 | $R_{wp}$ (%) | 11.0 |
| $R_I$ (%) | 2.47 | 2.88 | 4.87 | $R_I$ (%) | 4.84 |
| $R_{mag}$ (%) | 6.54 | 3.60 | 29.1 | $R_{mag}$ (%) | - |

**Table II.-** Main interatomic distances (Å) and angles (º) for ordered and disordered $Sr_2FeMoO_6$ perovskites

| | 15 K | | 500 K | | |
|---|---|---|---|---|---|
| Sample | Disordered | Ordered | Disordered | | Ordered |
| *I4/m* | | | | *Fm3m* | |
| $FeO_6$ octahedra | | | | $FeO_6$ octahedra | |
| Fe- O1 | 1.97(7) | 2.00(3) | 2.011(2) | Fe- O | 1.986(9) |
| Fe- O2 | 1.97(2) | 1.99(1) | 1.9508(6) | | |
| $MoO_6$ octahedra | | | | $MoO_6$ octahedra | |
| Mo- O1 | 1.98(7) | 1.95(3) | 1.947(2) | Mo- O | 1.967(9) |
| Mo- O2 | 1.98(2) | 1.95(1) | 2.0082(6) | | |
| Fe-O1-Mo | 180.0 | 180.0 | 180.0 | Fe-O-Mo | 180.0 |
| Fe-O2-Mo | 170.2(9) | 168.9(5) | 175.33(7) | | |